\documentstyle[11pt,cs10,html,epsf]{article}

\begin{document}

\title{An Inversion Technique to Derive Model Photospheres in Late-Type Stars
from High-Resolution Spectroscopy: The Sun}

\author{Carlos Allende Prieto, Basilio Ruiz Cobo, and Ram\'on J. Garc\'{\i}a
L\'opez}
\affil{Instituto de Astrof\'{\i}sica de Canarias}

\begin{abstract}

An inversion technique has been developed to recover LTE one-dimensional
model photospheres for late-type stars from very high-re\-so\-lu\-tion high
signal-to-noise stellar line  profiles. It is successfully applied to the Sun
using a set of unblended \ion{Ti}{1}, \ion{Ca}{1}, \ion{Cr}{1} and
\ion{Fe}{1} lines with accurate transition probabilities. Temperature
stratification, continuum flux, centre-to-limb variation and wings of strong
metal lines obtained from the resulting model are compared with those from
other well-known theoretical and empirical solar models and show the
reliability of the procedure.

\end{abstract}


\index{*Sun}

\section{The Inversion Code}

A new version of the inversion code developed by Ruiz Cobo \& del Toro
Iniesta (1992) to derive model photospheres from solar Stokes profiles has
been adapted to work with stellar flux line profiles as the only input.  The
program maximizes the agreement between the line spectrum emerging from the
star and a synthetic spectrum, modeling the temperature stratification of the
photosphere, the elemental chemical abundances, macroturbulence,
microturbulence and the projected rotational velocity. 

LTE and hydrostatic equilibrium constraints are imposed to the derived
photospheric structure. Starting from a given model photosphere, step by
step, the depth dependence of the temperature is allowed to be modified  in a
successively increasing number of nodes. Three snapshots  in the inversion
process are shown in Figure 1 (ordered a, b and c). The evolution of the
goodness of the line profile fitting from an isothermal  atmosphere to the
final result can be easily understood from the accompanying
\htmladdnormallink{animation}{file:peli.mpg}.

\begin{figure}
\plotfiddle{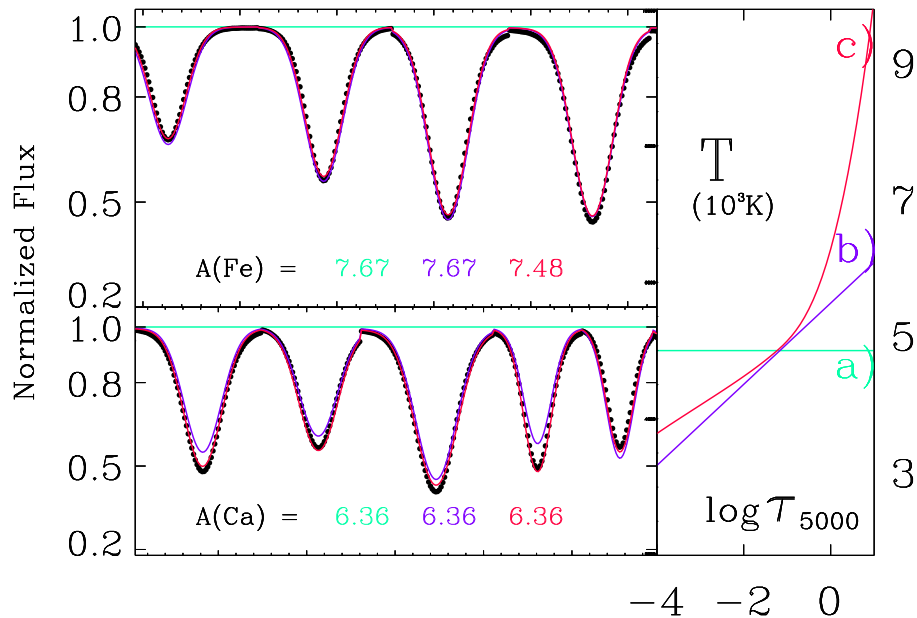}{7.5cm}{0}{65}{65}{-190}{-230}
\caption{Three steps in the inversion procedure. From a constant temperature
atmosphere, resulting in no spectral lines, to the final step where the line
profiles are well reproduced. Nine of the 40 lines employed for the inversion
are displayed.}
\end{figure}

The tests  performed have pointed out the possibility of recovering the
temperature in the photosphere from a few flux-calibrated line
profiles, but showed how small errors in the absolute flux calibration
translate in huge errors on the temperature scale. The possibility of
recovering the LTE photospheric structure by using a larger number of
locally normalized line profiles was also detected in the tests. This
means a remarkable advantage, because that information can be available
for any star, regardless of whether or not we know the distance to the star and
the errors in the spectrophotometry become larger than for the
solar case.

\section{It works for the Sun!}

The method is applied to the Sun, the best known cool star, where direct
comparison can be performed with other empirical and theoretical model
photospheres.  

The observations entering the inversion procedure are clean line
profiles from the solar flux spectrum of Kurucz et al. (1984), which is
available at \htmladdnormallinkfoot{the NOAO FTP site}
{ftp://pandora.tuc.noao.edu/fts}.  The input lines to be included in
the selection of solar lines by Meylan et al. (1993) were required, who
identified clean line wings in the same atlas by fitting Voigt
profiles. We also impose the condition for the line that its atomic
transition probability had been measured at the Oxford furnace (e.g.
Blackwell \&  Shallis 1979).

A total of 40 absorption lines of neutral iron, titanium, chromium and
calcium were entered into the inversion program. The clean  wings of the broad
\ion{Ca}{1}  $\lambda$6162 \AA , one of the few lines for which the
collisional width is accurately known, were also included.

\begin{figure}
\plotfiddle{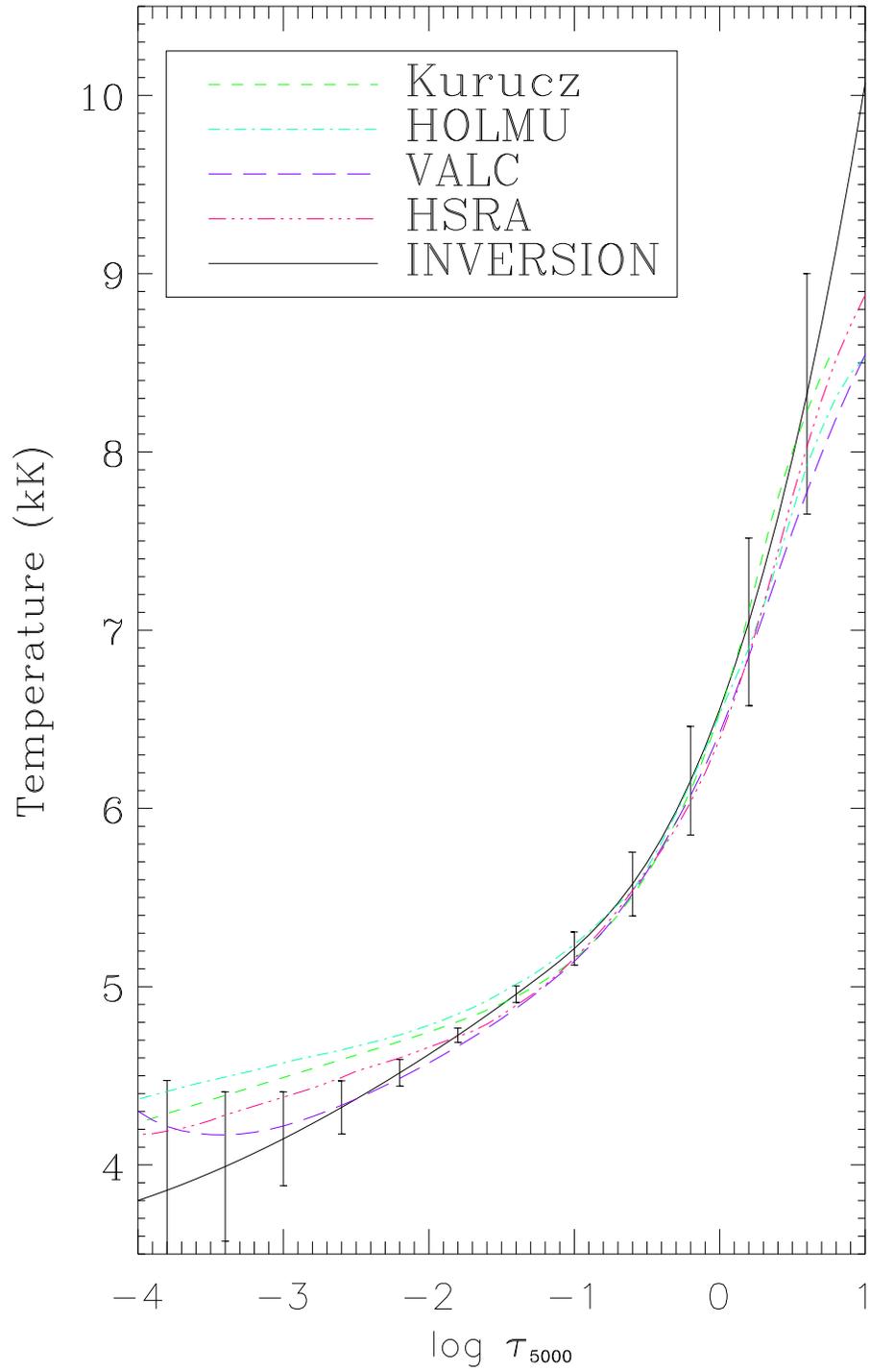}{17cm}{0}{80}{80}{-250}{-70}
\caption{The photospheric temperature structure for the Sun obtained from the
inversion method is directly compared with other classical models.}
\end{figure}

The method is able to extract information about the photospheric region
corresponding to the depth range where the employed lines are formed. The
derived  solar model photosphere  is shown in Figure 2. Following the
considerations by Meylan et al. (1993) to avoid blends, the whole line
profile or just one wing was accepted as input data. Agreement is found when
comparing its properties with observations and other well-known model
photospheres: 

\begin{itemize}
\item Limb darkening. The  continuum centre-to-limb variation predicted
by the INVERSION model keeps close to the observations for the entire optical
range (Figure 3).

\item Continuum.  Although the observed continuum cannot be easily
translated to the true one, which limits its usefulness in the absolute calibration of the line profiles, it is possible to correct the wide-band observations,
making use of high resolution spectra, to establish a lower limit to the true
continuum.   The pseudo-continuum deduced by Kurucz et al. (1984) from the data of Neckel
\& Labs (1984), is compatible with the predictions by HOLMU (Holweger 
\& M\"uller 1974),  Kurucz's model (Kurucz 1992) and the  photosphere from
the INVERSION.

\item Wings of strong lines. We have also checked that the spectral range
around  \ion{Ca}{1} $\lambda$ 6162 \AA\ is well reproduced by the model. Also,
the predicted wings of the Sodium D doublet, which are supposed to cover a
wide depth range of the photosphere, reproduce the solar measurements.

\end{itemize}

\begin{figure}
\plotone{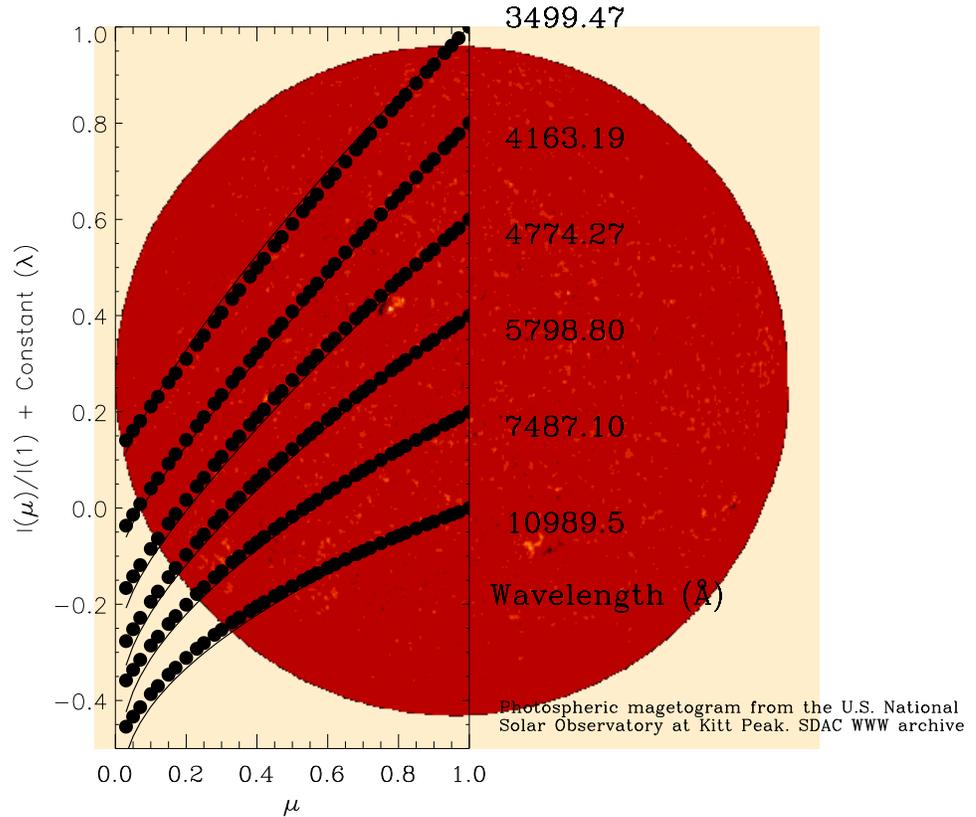}
\caption{The predicted specific intensity, normalized to that at the
disc-centre, is compared with the polynomial fits to the observations by Neckel
\& Labs (1994).}
\end{figure}

It is found that none of the input abundances (Anders \& Grevesse 1989) for
Cr, Ca, and Ti need to be changed to find the best fit of the observations,
but the Fe abundance is preferred to vary from the initially assumed  value
(7.67, in a scale where the hydrogen abundance is 12) to the lower meteoric
abundance 7.5

Assuming the rotation velocity (1.88 km s$^{-1}$), and the
microturbulence (0.6 km s$^{-1}$) as known, the code arrives at 1.7 km s$^{-1}$ for the  macroturbulent velocity.

\section{Summary}

From the inversion of {\sc normalized} line profiles we derive a
semi-empirical model photosphere for the Sun which reproduces the solar
continuum, the limb-darkening in the continuum, and the profiles of lines
which were not included in the input data for the inversion. 

To better fit the solar flux line spectrum would require the rejection of
the hypothesis of hydrostatic equilibrium, introducing multi-component models
or velocity fields giving place to asymmetries in the line profiles. Work for
the near future will point in this direction.

Making use of very high quality spectra (Allende Prieto et al. 1995) we plan
to apply this tool to metal-poor stars and compare the results with LTE
theoretical model atmospheres.

\acknowledgments

We sincerely thank H\'ector Socas Navarro for carrying out some calculations
for us. We are also grateful to Gabriel P\'erez D\'{\i}az, who helped us
with the movie work. NSO/Kitt Peak FTS data used here were produced by
NSF/NOAO.

\end{document}